\documentclass[12pt]{article}
\usepackage{amsfonts}

\begin{document}

\newtheorem{theorem}{Theorem}
\newtheorem{lemma}{Lemma}
\newtheorem{proposition}{Proposition}
\newtheorem{definition}{Definition}

%     1mdemeixnr140808.tex ex 1mdemeixnr0106.tex
\centerline{\bf \Large Fermionic Meixner probability distributions, }
\centerline{\bf \Large Lie algebras and
  quadratic Hamiltonians }

\bigskip\bigskip

\centerline{\Large \bf L. Accardi (1), I.Ya. Aref'eva, I.V.Volovich (2)}

\bigskip

\begin{center}
(1) Centro Vito Volterra\\
Universit\`a di Roma ``Tor Vergata"\\
Roma I-00133 Italy
\end{center}
\begin{center}
(2) Steklov Mathematical Institute,\\
 Gubkin St.8, GSP-1, 117966, Moscow, Russia,  \\
volovich@mi.ras.ru, arefeva@mi.ras.ru
\end{center}

\tableofcontents

\eject

\textbf{Abstract}
We introduce the quadratic Fermi algebra, which is a Lie algebra,
and show that the vacuum distributions of the associated Hamiltonians
define the fermionic Meixner probability distributions.
In order to emphasize
the difference with the Bose case, we apply a modification
of the method used in the above calculation to obtain a simple and straightforward
classification of the $1$--dimensional Meixner laws in terms of homogeneous
quadratic expressions in the Bose creation and annihilation operators.
There is a huge literature of the Meixner laws but this, purely quantum
probabilistic, derivation seems to be new.
Finally we briefly discuss the possible multi--dimensional
extensions of the above results.

\section{Introduction}

The program of {\bf quadratic quantization} was introduced in the paper
\cite{[AcLuVo99]} where the realizability of this program was proved by
explicit construction of the Fock representation of the quadratic
Bose algebra (see Definition (\ref{df-slB(2,C)}) below).
In the paper \cite{[AcFrSk00]} the current algebra of the quadratic
Bose algebra over $\mathbb R^d$ was identified with the corresponding
current algebra of $sl(2,\mathbb R)$. This allowed to include the
problem of constructing representations of the quadratic Bose algebra
into the general theory of factorizable representations of Lie algebras
(see \cite{[PaSch72]} and the bibliography therein) and its extension to
quantum Levy processes on $*$--bi--algebras
(see \cite{[AcSchuvW87]}, \cite{[Schu93]}).
Combining these techniques with the fact that the
irreducible unitary representations of $sl(2,\mathbb R)$ are classified,
in \cite{[AcFrSk00]} a class of unitary representation of the
quadratic Bose algebra was built and the vacuum distributions of the
generalized field operators were identified with the $3$ non standard
(i.e. non Gauss or Poisson) Meixner probability distributions.\\

The problem of the existence of a Fermi analogue of the
quadratic Bose algebra was investigated in  \cite{[AcArVo14a]} where:\\
(i) a natural candidate for the role of {\bf quadratic Fermi algebra}
was constructed (see Definition (\ref{df-slF(2,C)}) below);\\
(ii) the Lie algebra isomorphism between the quadratic Fermi algebra and
$sl(2,\mathbb R)$ was proved (this in particular implies the Lie
algebra isomorphism between the quadratic Fermi and Bose algebras);\\
(iii) it was shown that the quadratic Fermi and Bose algebras cannot
be\\ $*$--isomprphic for the natural involutions induced by their
explicit constructions in terms of Fermions or Bosons respectively.\\

The non $*$--isomprphism result, mentioned in item (iii) above,
naturally rises the
problem of identifying the vacuum distributions of the field operators
in the Fock representation of the quadratic Fermi algebra. This problem is
solved in section \ref{Vac-exp-Fer-quad-Ham} of the present paper.
From this one
can deduce a weaker form of the above mentioned non
$*$--isomprphism result, namely that there cannot exist a $*$--isomprphism
between the two algebras mapping the Fermi quadratic annihilator into
a multiple of its Bose analogue. In fact, if such a $*$--isomprphism existed,
then the restriction of the Fock states of the two representations to
the Cartan sub--algebras of the two algebras should give rise to the
same class of probability measures. However, comparing the results
of section \ref{Vac-exp-Fer-quad-Ham} below with those of \cite{[AcFrSk00]}
(see also section \ref{1-mde-Bos-quad-Ham} below)
one immediately verifies that this is not the case.\\

In section \ref{1-mde-Bos-quad-Ham} below it is shown that the
method developed for the calculation
of the Fermi vacuum distributions can be applied to the Bose case
leading to a very simple and elegant classification of all possible
vacuum distributions of homogeneous quadratic Boson
Hamiltonians in terms of Meixner laws. On both of these
topics, {\bf separately}, there exists a vast literature  \cite{[Friedr51]},
\cite{Berezin} for vacuum distributions of quadratic expressions of
usual Boson fields (where however the connection with Meixner classes
was missing) and \cite{GRIGa}, \cite{GRIGc}
for the probabilistic aspects of Meixner laws and the connection
between the two was established in  \cite{[AcFrSk00]} using the theory
of orthogonal polynomials. The method used here is different and we have
reasons to believe that it can be extended to the multi--dimensional case
(in this case the orthogonal polynomial method can only be applied to
products of Meixner measures).  The problems related with the multi--dimensional
case are briefly outlined in the final section \ref{n-dim-cse} .\\

We emphasize that the present paper is only the beginning of our
investigation the quadratic Fermi algebra.
In this direction there are many deep problems which have not yet an
answer. Among them we mention:\\
(I) the existence of the Fock representation
the current algebra over $\mathbb R^d$ of the quadratic Fermi algebra
(for this the methods of  \cite{[AcFrSk00]} should be sufficient); \\
(II) in case of existence, the vacuum distribution of the corresponding fields;\\
(III) the validity, also in the Fermi case, of the no--go theorems
proved in \cite{[SNIA99]}, \cite{[AcFrSk00]}, \cite{[AcBouFr06_RPQWN]},
\cite{[AcBou06_HP_q-Deformed_WN]}.\\
These problems are now under investigation.\\

Finally let us mention that, after completing the present paper, we
received the very interesting preprint \cite{[PopaSta14]} including
a different approach to the notion of multi--dimensional Meixner
random variables, based on the quantum decomposition of a classical
multi--dimensional random variable with all moments and on its
characterization in terms of commutators (see \cite{[AcKuoSta07]}).
This naturally poses the problem to clarify the mutual relationships
between these two approaches.

\subsection{Triviality of the $1$--mode quadratic Fermi algebra}
\label{Triv-1-mde-quadr-F-alg}

Recall that the $1$--mode CAR algebra, denoted $CAR(1)$,
is the $*$--algebra with identity $1$, generators $\{a,a^+,1\}$
called the {\bf Fermi creation and annihilation operators} and relations
$$
(a^+)^*=a
$$
\begin{equation}\label{CAR1}
\{a,a\}=\{a^+,a^+\}=0 \qquad ;\qquad
\{a,a^+\}:=aa^+ + a^+a=1
\end{equation}
As a consequence of (\ref{CAR1}), one has
$$
a^{2}=a^{+2}= 0
$$
which implies that both $a^{+}a$ and $aa^{+}$
are orthogonal projections.
In fact they  are clearly self--adjoint and:
Because of (\ref{CAR1}) the pair
$\{aa^{+},a^{+}a\}$ is a partition of the identity.
Thus, for a $1$--mode representation,
the quadratic algebra
$$
\{a^2,a^{+2}, aa^+\}=\{0,aa^+\}
$$
is generated by the single projection, hence it is abelian.

\subsection{The quadratic Fermi algebra}
\label{quadr-Ferm-alg}

A possible candidate for the role of (non--trivial) quadratic Fermi
algebra is constructed as follows.
Consider the $2$--mode CAR algebra, denoted $CAR(2)$,
i.e. the associative $*$--algebra with identity $1_F$, generators:
$a_i,a_j^+$ ($i,j=1,2$) and relations
\begin{equation}\label{CAR}
a_j=(a_j^+)^*
\quad;\quad
\{a_i,a_j\}=\{a_i^+,a_j^+\}=0\quad;\quad i,j=1,2
\end{equation}
$$
\{a_i,a^+_j\}:=a_ia^+_j+a^+_ja_i=:[a_i,a^+_j]_+=\delta_{ij}\cdot 1_F
$$
\begin{lemma}\label{lm-quadr-CAR}{\rm
In the above notations, defining
$$
F^{-}:=a_1a_2
$$
$$
F^{+}:=a^+_2a^+_1 =(F^{-})^*
$$
$$
N_F=N_F^*:=a^+_2a_2+a^+_1a_1
$$
the following {\bf quadratic Fermi commutation relations} hold:
\begin{equation}\label{CAR2a}
[F^{-},F^{+}]=-N_F+1_F=:M_F
\end{equation}
\begin{equation}\label{CAR2b}
[N_F,F^{+}]=[M_F,F^{+}]=2F^{+}
\end{equation}
\begin{equation}\label{CAR2c}
[N_F,F^{-}]=[M_F,F^{-}]=-2F^{-}
\end{equation}
\begin{equation}\label{CAR2d}
[X,1_F]=0 \qquad ;\qquad X=F^{\pm} \ , \
\end{equation}
}\end{lemma}
{\bf Remark}
Notice that $N_F\neq 1_F$, in fact for example
$$
a^+_1N_F=a^+_1(a^+_2a_2+a^+_1a_1)=a^+_1a^+_2a_2
=a^+_2a_2a^+_1\neq N_Fa^+_1
$$
Therefore the right hand side of (\ref{CAR2a}) is not equal to zero.
Denoting
$$
S_1:=F^{+}\quad ;\quad S_2:=F^{-}
\quad ;\quad S_3:=N_F \quad ;\quad S_0:=1_F
$$
(\ref{CAR2a}), (\ref{CAR2b}), (\ref{CAR2c})
become equivalent to
\begin{equation}\label{CAR2aS}
[S_1,S_2]=-S_3+S_0 \quad ;\quad
[S_3,S_1]=-2S_1    \quad ;\quad
[S_3,S_2]=2S_2
\end{equation}
$$
[S_0,X]=0
$$
\begin{equation}\label{CAR2dS}
(S_1)^*=S_2  \quad ;\quad
(S_3)^*=S_3 \quad ;\quad
(S_0)^*=S_0
\end{equation}

\begin{definition}\label{df-slF(2,C)}{\rm
 $*$--Lie algebra with generators
$(S_1,S_2,S_3,S_0)$ and relations (\ref{CAR2aS}),  (\ref{CAR2dS})
is called the quadratic Fermi algebra.
}\end{definition}

We will consider a representation of the quadratic Fermi algebra
in a Hilbert space with the vacuum vector $\psi_0$ such that
$a_j\psi_0 =0, j=1,2.$

\subsection{Vacuum expectations of quadratic Fermi Hamiltonians}
\label{Vac-exp-Fer-quad-Ham}

The following problem naturally arises:
{\it Which are the vacuum distributions of the
field operators of the quadratic Fermi algebra?}\\
To discuss this problem we define, for $\alpha,\beta \in \mathbb{R}$,
the homogenous quadratic Hamiltonian:
\begin{equation}\label{ham1}
H=\alpha(a^+_1a_1+a^+_2a_2)+\beta(a^+_1a^+_2+a_2a_1)\label{ham1}
\end{equation}
and the associated time evolution, uniquely determined, on the
polynomial $*$--algebra generated by the $a_i, a_i^+$ ($i=1,2$),
as the solutions of the equations
\begin{equation}\label{ham-ev1}
\partial a_j(t):= i[H,a_j(t)] \qquad ; \qquad a_j(0)=a_j \ , \ j=1,2
\end{equation}
\begin{lemma}\label{lemma9}{\rm
Suppose $\alpha$, $\beta$ are arbitrary real numbers, such that
\begin{equation}\label{defomega2}
\omega=\sqrt{\alpha^2+\beta^2}>0
\end{equation}
Then the Heisenberg evolutions of
$a_i, a_i^+$, $i=1,2$ are given by:
\begin{equation}
\label{e101}
a_1(t)=\left(\cos\omega t-{i\alpha\over\omega}\,\sin\omega t\right)a_1-
{i\beta\over\omega}\,\sin\omega t\cdot a_2^+
\end{equation}
\begin{equation}
\label{e102}
a_2(t) = \left(\cos \omega t -
{i\alpha \over {\omega}}\,\sin{\omega} t\right)
a_2 + {i\beta \over {\omega}}\,\sin {\omega} t\cdot a^+_1
\end{equation}
\begin{equation}
\label{e103}
a_1^+(t) = \left(\cos \omega t + \frac{i \alpha}{\omega} \sin \omega t \right) a_1^+ +
\frac{i\beta}{\omega} \sin \omega t \cdot a_2
\end{equation}
\begin{equation}
\label{e104}
a_2^+(t) =
\left(\cos \omega t + \frac{i\alpha}{\omega} \sin \omega t\right)a_2^{+} -
\frac{i \beta}{\omega} \sin \omega t \cdot a_1
\end{equation}
}\end{lemma}
\textbf{Proof.}
Note that
$$
i[H,a_1]=i[\alpha a^+_1a_1+\beta a^+_1a^+_2,a_1]=
$$
\begin{equation}
=i\alpha a^+_1a_1a_1+i\beta a^+_1a^+_2a_1-
 i\alpha a_1a^+_1a_1-i\beta a_1a^+_1a^+_2=\label{a1}
\end{equation}
$$=-i\beta a^+_1a_1a^+_2-i\alpha a_1+i\beta a^+_1a_1a^+_2-i\beta a^+_2=-i\alpha a_1-i\beta a^+_2$$
$$i[H,a_2]=i[\alpha a^+_2a_2+\beta a^+_1a^+_2,a_2]=$$
\begin{equation}
=i\alpha a^+_2a_2a_2+i\beta a^+_1a^+_2a_2-i\alpha a_2a^+_2a_2-i\beta a_2a^+_1a^+_2=\label{a2}
\end{equation}
$$=i\beta a^+_1a^+_2a_2-i\alpha a_2-i\beta a^+_1a^+_2a^+_2a_2+i\beta a^+_1=-i\alpha a_2+i\beta a^+_1$$

%\end{document}

Now note that
$$
\frac{da_1(t)}{dt} = i[H,a_1(t)] = i[H,e^{itH} a_1 e^{-itH}] =
e^{itH} i[H,a_1] e^{-itH} = $$ $$= e^{itH}\left(-i \alpha a_1 -i \beta a_2^+ \right) e^{-itH} =
-i \alpha a_1(t) - i\beta a_2^+(t)
$$
Similarly,
$$
\frac{da_2(t)}{dt} = -i\alpha a_2(t)+i\beta a^+_1(t)
$$
It is easy to deduce the equations for
$\frac{da_1^+(t)}{dt}$ and $\frac{da_2^+(t)}{dt}$.
The result is summarized in the following equation:
\begin{equation}\label{def-J}
\frac{d}{dt}
\left(
\begin{array}{c}
a_1(t) \\
a_2(t) \\
a_1^+(t) \\
a_2^+(t) \\
\end{array}
\right)
=
i
\left(
\begin{array}{cccc}
-\alpha    &    0       &     0   & -\beta  \\
0          & -\alpha    & \beta   &   0     \\
0          & \beta      & \alpha  &   0     \\
-\beta     &  0         &   0     &  \alpha \\
\end{array}
\right)
%\end{document}
\left(
\begin{array}{c}
a_1(t) \\
a_2(t) \\
a_1^+(t) \\
a_2^+(t) \\
\end{array}
\right)
=: iJ\left(
\begin{array}{c}
a_1(t) \\
a_2(t) \\
a_1^+(t) \\
a_2^+(t) \\
\end{array}
\right)
\end{equation}

%\end{document}
with initial condition $a_j^\pm(0)=a_j^\pm$. Therefore, the solution
of this differential equation is:
$$
\left(
\begin{array}{c}
a_1(t) \\
a_2(t) \\
a_1^+(t) \\
a_2^+(t) \\
\end{array}
\right)=
\exp \left( it\left(
\begin{array}{cccc}
-\alpha    &    0       &     0   & -\beta  \\
0          & -\alpha    & \beta   &   0     \\
0          & \beta      & \alpha  &   0     \\
-\beta     &  0         &   0     &  \alpha \\
\end{array}
\right)\right)
\left(
\begin{array}{c}
a_1 \\
a_2 \\
a_1^+ \\
a_2^+ \\
\end{array}
\right)
%\end{document}
$$
In the notation (\ref{def-J}) $J^2$ is equal to
$$
\pmatrix{ -\alpha    &    0       &     0   & -\beta  \cr
					0          & -\alpha    & \beta   &   0     \cr
					0          & \beta      & \alpha  &   0     \cr
					-\beta     &  0         &   0     &  \alpha \cr}
					\pmatrix{ -\alpha    &    0       &     0   & -\beta  \cr
					0          & -\alpha    & \beta   &   0     \cr
					0          & \beta      & \alpha  &   0     \cr
					-\beta     &  0         &   0     &  \alpha \cr}
=\pmatrix{ \alpha^2+\beta^2    &    0       &     0   & 0  \cr
					0          & \alpha^2+\beta^2    & 0   &   0     \cr
					0          &   0    & \alpha^2+\beta^2  &   0     \cr
					0    &  0         &   0     &  \alpha^2+\beta^2 \cr}
$$
$$
= (\alpha^2+\beta^2)\cdot 1=\omega^2\cdot 1
$$
Therefore
$$
J^{2n+1}= \omega^{2n} J \qquad ;\qquad J^{2n}= \omega^{2n}
$$
and from this one verifies that the matrix exponential is
$$
\exp \left( itJ \right)
=
\left(%
\begin{array}{cccc}
  \cos \omega t - \frac{i\alpha}{\omega} \sin \omega t & 0 & 0 & -\frac{i\beta}{\omega}\sin \omega t \\
  0 & \cos \omega t - \frac{i\alpha}{\omega} \sin \omega t & \frac{i\beta}{\omega}\sin \omega t  & 0 \\
  0 & \frac{i\beta}{\omega}\sin \omega t  & \cos \omega t + \frac{i\alpha}{\omega} \sin \omega t & 0 \\
  -\frac{i\beta}{\omega}\sin \omega t  & 0 & 0 & \cos \omega t + \frac{i\alpha}{\omega} \sin \omega t \\
\end{array}%
\right)
$$

%\end{document}

Hence, we get Eqs. (\ref{e101}-\ref{e104}).
\begin{lemma}
\label{lemma10}
Denote
\begin{equation}
\label{e10_01}
f(t):=\langle\psi_0,e^{itH}\psi_0\rangle
\end{equation}
Then $f$ satisfies the equation
\begin{equation}
\label{e10_02}
\left(1-\frac{\beta^2}{ \omega^2} \sin^2 \omega t\right)\frac{df}{dt}
=
-\frac{\beta^2}{2\omega} \sin  \omega t \left(\cos {\omega} t + \frac{i\alpha}{\omega} \sin \omega t\right)f
\end{equation}
with $\omega$ given by (\ref{defomega2}) and
\begin{equation}
\label{e10_021}
f(0)=1
\end{equation}
\end{lemma}
\textbf{Proof.}
Let $f(t)$ be given by (\ref{e10_01}).
We have
$$
{df\over dt}\,=
i\beta \langle\psi_0,e^{itH}a_1^+ a_2^+\psi_0\rangle=i\beta\langle\psi_0,a_2 a_1 e^{itH}\psi_0\rangle
$$
Let us use the identity:
\begin{equation}
\label{e10_025}
e^{itH} a_1^{+} a_2^{+} = e^{itH} a_1^+ I a_2^+ I = e^{itH} a_1^+  e^{-itH} e^{itH} a_2^+ e^{-itH} e^{itH}
= a_1^+(t) a_2^+(t) e^{itH}
\end{equation}
Substituting $a_1^+(t)$ $a_2^+(t)$ from Lemma \ref{lemma9}, we get:
\begin{equation}
\label{e10_03}
\frac{df}{dt} = i\beta \langle\psi_0,a_1^+(t) a_2^+(t) e^{itH}\psi_0\rangle =
\end{equation}

%\end{document}

$$
=
i \beta
\left\langle \psi_0,
\frac{i\beta}{\omega} \sin  \omega t \cdot a_2
\left(\left(\cos {\omega} t + \frac{i\alpha}{\omega} \sin \omega t\right) a_2^+ -
\frac{i\beta}{\omega} \sin \omega t \cdot a_1\right)e^{itH}\psi_0\right\rangle
$$

%\end{document}

$$
=-\frac{\beta^2}{\omega}\sin  \omega t \left(\cos {\omega} t + \frac{i\alpha}{ \omega} \sin  \omega t\right)\langle\psi_0,a_2a_2^+e^{itH}\psi_0\rangle
$$
$$
+i\beta\frac{\beta^2}{\omega^2}\sin^2  \omega t \langle\psi_0,a_2a_1e^{itH}\psi_0\rangle
$$

%\end{document}

Since
$$
\langle \psi_0, a_2a_2^+ e^{itH}\psi_0 \rangle = \langle \psi_0, (1-a_2^+a_2) e^{itH}\psi_0 \rangle =
\langle \psi_0,  e^{itH}\psi_0 \rangle = f(t)
$$
and according to (\ref{e10_03}),
$$
\langle \psi_0, a_2 a_1 e^{itH}\psi_0 \rangle = \frac{1}{i\beta} \frac{df}{dt}
$$

%\end{document}

we get:
$$
\frac{df}{dt} = -\frac{\beta^2}{\omega}\sin  \omega t \left(\cos {\omega} t + \frac{i\alpha}{\omega} \sin \omega  t\right)f
+\frac{\beta^2}{\omega^2}\sin^2 \omega t \frac{df}{dt}
$$
which is (\ref{e10_02}). The initial condition (\ref{df0}) is obvious.
\begin{lemma}
\label{lemma11}
\begin{equation}
\label{e11_01}
\langle\psi_0,e^{itH}\psi_0\rangle =
\left(\cos \omega t - \frac{i\alpha}{\omega}\sin\omega t\right) e^{i\alpha t}
\end{equation}
\end{lemma}
{\bf Proof}. By direct calculations one can check that the function
(\ref{e11_01}) is the solution of the differential equation (\ref{e10_02}) with the
initial condition (\ref{e10_021}).

\begin{theorem}
The vacuum expectation
${f(t)=\langle\psi_0,e^{itH}\psi_0\rangle}$ is the characteristic
function of the Bernoulli random variable $X$  with distribution
\begin{equation}\label{pX}
p_X(x)=\frac{1}{2}\left(\left(1-\frac{\alpha}{\omega}\right)
\delta(x+\alpha+\omega)+
\left(1+\frac{\alpha}{\omega}\right)\delta(x+\alpha-\omega)\right)
\end{equation}
\end{theorem}

%\end{document}

\textbf{Proof.}
This follows from Lemma \ref{lemma11}. In fact:
$$
F^{-1}\left[\left(\cos \omega t - \frac{i\alpha}{\omega}\sin\omega t\right) e^{i\alpha t}\right]=
F^{-1}\left[\left(\frac{1}{2}\left(e^{i\omega t}+ e^{-i\omega t}\right) - \frac{\alpha}{2\omega}\left(e^{i\omega t}-e^{-i\omega t}\right)\right) e^{i\alpha t}\right]
$$

%\end{document}

$$
=\frac{1}{2}F^{-1}\left[\left(1-\frac{\alpha}{\omega}\right)e^{i(\omega+\alpha) t}+
\left(1+\frac{\alpha}{\omega}\right)e^{i(\alpha-\omega) t}\right]=
$$
%\end{document}

$$
=\frac{1}{2}\left(\left(1-\frac{\alpha}{\omega}\right)\delta(x+\alpha+\omega)+
\left(1+\frac{\alpha}{\omega}\right)\delta(x+\alpha-\omega)\right).
$$

{\bf Remark}. Notice that the limit of (\ref{pX}) as $\omega\to 0$
in such a way that $\alpha/\omega\to c\in\mathbb R$ exists and
(independently of $c$) is equal to
\begin{equation}\label{p0X}
p_{0,X}(x)=\delta(x)
\end{equation}
\begin{definition}\label{Fer-Meix-cl}
For varying $\alpha$ and $\beta$ the family (\ref{pX}) can be easily
seen to coincide with the class of all Bernoulli distributions.
The class correspondig to $\alpha \neq \beta$ will be called the
$1$--st Fermionic Meixner classes. The $2$--d Fermionic Meixner class,
correspondig to $\alpha=\beta$, consists of the single
$\delta$--measure (\ref{p0X}).
\end{definition}

%\end{document}

\section{$1$--mode Bose quadratic Hamiltonians}

\label{1-mde-Bos-quad-Ham}

In this section we recall some known facts about quadratic $1$--mode Hamiltonians.
Let $\{ {\cal H}, a^+, a , \psi_0 \} $ be the (unique up to unitary equivalence)
Fock representation of the $1$--mode canonical commutation relations (CCR).
This means that ${\cal H}$ is a Hilbert space, with scalar product denoted
by $\langle\cdot,\cdot\rangle$, $\psi_0\in{\cal H}$ is a unit vector, called
the vacuum, and $a^+$, $a$ are operators on ${\cal H}$, called respectively
creation and annihilation operators satisfying the Fock relation
$$
a\psi_0=0
$$
and such that, for each $n\in\mathbb{N}$, $\psi_0$ is in the domain of $(a^+)^n$,
the linear span of the vectors $(a^+)^n\psi_0$ is dense in ${\cal H}$ (in this case we say
that $\psi_0$ is cyclic for the polynomial algebra generated by $a^+$ and $a$)
and on this domain the $1$--mode CCR
\begin{equation}\label{e1}
[a,a^+]=1
\end{equation}
are satisfied.
\begin{definition}\label{df-slB(2,C)}{\rm
The $4$--dimensional complex $*$--Lie algebra with generators
$(T_1,T_2,T_3,T_0)$ and relations
\begin{equation}\label{CCR1aT}
[T_2,T_1]=T_3  \quad ;\quad
[T_3,T_1]=2T_1  \quad ;\quad
[T_3,T_2]=-2T_2\quad ;\quad
[T_0,X]=0
\end{equation}

%\end{document}

\begin{equation}\label{CCR2dT}
(T_1)^*=T_2\quad ;\quad
(T_3)^*=T_3\quad ;\quad (T_0)^*=T_0
\end{equation}
is called the quadratic Bose algebra
}\end{definition}

%\end{document}

{\bf Remark}. In the notation (\ref{e1}, the identifications:
$$
T_1:=a^{+2} \quad ;\quad  T_2:=a^{2} \quad ;\quad  T_3:=4a^{+}a+2
$$
give a concrete realization of the quadratic Bose algebra
on the space of the Fock representation of the Heisenberg algebra.
Note the difference with Schwinger bosons. Schwinger uses a doublet  of bosons
(see for example \cite{Schwi82}, \cite{Anish-Mathur-Raych09})
while we use a singlet.

The most general symmetric, real homogeneous, quadratic
expression in the variables $a^+ ,a $ has the form:
\begin{equation}\label{e2}
H :=\frac{1}{2}\alpha\left({a^+}\right)^2
+\frac{1}{2}\bar{\alpha}a^2 + \beta a^+ a
\end{equation}
where $\alpha \in {\Bbb C}$, $\beta \in {\mathbb{R}}$. $H$ is not identically
zero if and only if
\begin{equation}\label{nontriv}
|\alpha|^2+\beta^2>0
\end{equation}
Introducing the matrix
$$
h:=\pmatrix{\alpha & \beta \cr
\beta  & \bar{\alpha} \cr}
$$
$H$ can be represented, up to an inessential additive constant, in the form:
$$
H=\frac{1}{2} + \frac{1}{2} (a^+,a) h\pmatrix{a^+     \cr  a\cr}
$$
Without loss of generality it can be assumed that
\begin{equation}\label{e2real}
\alpha\in\mathbb{R}_+
\end{equation}
in fact if $\alpha = |\alpha| e^{i\theta}$, where $\theta\in \mathbb{R}$,
then the gauge transformation
$$
a \to \tilde{a} := e^{-i\theta/2}a
\qquad ;\qquad  a^+ \to \tilde{a}^+ := e^{i\theta/2}a^+
$$
leaves invariant the commutation relations (\ref{e1}) and the quadratic
Hamiltonian $H$ in terms of $\tilde{a}$ and $\tilde{a}^+$ becomes:
$$
H=\frac{1}{2}|\alpha|\left(\left({\tilde{a}^+}\right)^2
+\tilde{a}^2 \right) + \beta \tilde{a}^+ \tilde{a}
$$
$H$ generates the $1$--parameter unitary group $U_t=e^{itH}$ and the Heisenberg
evolution of any operator $A$ under this group (we can simply say under $H$)
is defined by
\begin{equation}\label{heisev-h}
A(t):=e^{itH} A e^{-itH}
\end{equation}
\begin{lemma}\label{lemma2}{\rm
Let $\alpha, \beta \in \mathbb{R}$ satisfy (\ref{nontriv})
and let $\omega$ denote the positive square root of
\begin{equation}\label{defomega}
\omega^2:=\beta^2-\alpha^2=-\det h
\end{equation}
Then, if $\omega  \ne 0$ the Heisenberg evolutions of $a, a^+$ under $H$ are given by:
\begin{equation}\label{evano0}
a(t)=\left(\cos\omega t-{i\beta\over\omega}\,\sin\omega t\right)a-
{i\alpha\over\omega}\,\sin\omega t\cdot a^+
\end{equation}
\begin{equation}\label{eva+no0}
a^+(t)=  \left(\cos\omega  t+
{i\beta\over\omega }\,\sin\omega  t\right)
a^++{i\alpha\over\omega }\,\sin\omega  t\cdot a
\end{equation}
If $\omega  = 0$ they are given by the limit of the expressions (\ref{evano0})
and (\ref{eva+no0}) as $\omega  \to 0$, i.e.:
\begin{equation}\label{eva0}
a(t)=\left(1-i\beta t\right)a - i\alpha t a^+
\end{equation}
\begin{equation}\label{eva+0}
a^+(t)=  \left(1+i\beta t\right)a + i\alpha t a^+
\end{equation}
}\end{lemma}
\noindent{\bf Remark}.
Notice that the evolution (\ref{evano0}), (\ref{eva+no0}), hence also
(\ref{eva0}), (\ref{eva+0}), are invariant under change of sign of $\omega$.
In particular they do not depend on the choice of the real square
root of $\omega$.\\
\noindent{\bf Proof}.
Note that
$$
\frac{da(t)}{dt} = i[H,a(t)] = i[H,e^{itH} a e^{itH}]
= ie^{itH}[H,a]e^{-itH}=
$$
\begin{equation}\label{eqa1}
 = ie^{itH}\left(-\beta a -\alpha a^+\right) e^{-itH} =
-i\beta a(t)-i\alpha a^+(t)
\end{equation}
Similarly,
\begin{equation}
{da^+(t)\over dt}\,= i\beta a^+(t) + i \alpha a(t) \label{eqa+1}
\end{equation}
With the notations $a_t \equiv a(t)$, $\dot a_t \equiv da(t)/dt$, from
(\ref{eqa1}) we get
\begin{equation}
i\alpha a_t^+=-i\beta a_t-\dot a_t
\label{eqa+2}
\end{equation}
or equivlently, for
$$
\alpha \neq 0
$$
\begin{equation}
a_t^+=-{\beta\over\alpha}\,a_t-{1\over i\alpha}\,\dot a_t
\end{equation}
Taking the time derivative of this equation and
using (\ref{eqa+1}), (\ref{eqa+2}) we get

%\end{document}

\begin{equation}
-{\beta \over \alpha}\,\dot a-{1 \over i\alpha}\,\ddot a =\dot a^+
=
i\beta a^+ + i\alpha a
=
-{i\beta^2\over \alpha}\,a-{\beta\over  \alpha}\,\dot a+ i\alpha a
\end{equation}
\begin{equation}
-{1\over i\alpha}\,\ddot a={\beta^2\over i\alpha}\,a+i\alpha a
\end{equation}
Therefore
\begin{equation}
\ddot a+(\beta^2-\alpha^2)a=0\label{eqaa}
\end{equation}
In the notation (\ref{defomega}) we write equation (\ref{eqaa}) in the form
\begin{equation}
\label{e3}
\ddot a+\omega^2a=0
\end{equation}
Since $U_0=1$, $a(0)=a$ and $a^+(0)=a^+$. From (\ref{eqa1}) we get
the initial conditions:
\begin{equation}
\label{e4}
a(0)=a\ ,\quad\dot a(0)=-i\beta a-i \alpha a^+
\end{equation}
Thus, $a(t)$ is the solution of Eq. (\ref{e3}) with the initial
conditions (\ref{e4}). Looking for a solution of he form
$$
a(t)=A\cos\omega t+B\sin\omega t
$$
we find
$$
a=A\ ;\quad-i\beta a-i\alpha a^+=\omega B$$
so that
$$
a(t)=
a\cos\omega t-{i\beta\over\omega}\,a\sin\omega t-
{i\alpha\over\omega}\,a^+\sin\omega t
$$
$$
=\left(\cos\omega t-{i\beta\over\omega}\,\sin\omega t\right)\cdot a-
{i\alpha\over\omega}\,\sin\omega t\cdot a^+
$$
This proves (\ref{evano0}). By taking the adjoint we get (\ref{eva+no0}).

%\end{document}

Finally it is clear that (\ref{eva0}) and (\ref{eva+0}) are the limits of
(\ref{evano0}), (\ref{eva+no0}) respectively  as $\omega  \to 0$ and that
(\ref{eva+0}) satisfies equation (\ref{eqa+1}) with initial condition $a^+(0)=a^+$.

\begin{lemma}\label{lemma3}
Denote
\begin{equation}
f(t):=\langle\psi_0,e^{itH}\psi_0\rangle\label{dfchfu}
\end{equation}
If $\omega \ne 0$, then $f$ satisfies the equation
\begin{equation}
\label{eq6}
\left(1+\frac{\alpha^2}{\omega ^2} \sin^2\omega  t\right)\frac{df}{dt}
=
-\frac{\alpha^2}{2\omega } \sin \omega  t \left(\cos \omega  t +
\frac{i\beta}{\omega } \sin \omega  t\right)f
\end{equation}

with $\omega$ given by (\ref{defomega}) and
\begin{equation}
f(0)=1\label{df0}
\end{equation}
\end{lemma}

%\end{document}

\noindent{\bf Proof}. Let $f(t)$ be given by (\ref{dfchfu}). We have
$$
{df\over dt}\,=  \frac{i\alpha}{2}\langle\psi_0,e^{itH}a^{+2}\psi_0\rangle=
\frac{i\alpha}{2}\langle\psi_0,a^2e^{itH}\psi_0\rangle
$$
Using the identity:
\begin{equation}\label{e5}
e^{itH} (a^{+})^2 =
e^{itH} a^+  e^{-itH} e^{itH} a^+ e^{-itH} e^{itH}= a^+(t) a^+(t) e^{itH}
\end{equation}
and substituting $a^+(t)$ from Lemma (\ref{lemma2}), we get:
$$
\frac{df}{dt} = \frac{i\alpha}{2}\langle\psi_0,a^+(t) a^+(t) e^{itH}\psi_0\rangle =
$$

%\end{document}

$$
=
\frac{i\alpha}{2}
\langle \psi_0,\frac{i\alpha}{\omega } \sin \omega  t \cdot a
((\cos \omega  t +
\frac{i\beta}{\omega } \sin \omega  t) a^+
$$
$$
+
\frac{i\alpha}{\omega }
\sin \omega  t \cdot a)e^{itH}\psi_0\rangle
$$

%\end{document}

$$
=-\frac{\alpha^2}{2\omega }\sin \omega  t \left(\cos \omega  t +
\frac{i\beta}{\omega } \sin \omega  t\right)
\langle\psi_0,aa^+e^{itH}\psi_0\rangle
$$
$$
-\frac{i\alpha^3}{2 \omega ^2}\sin^2 \omega  t
\langle\psi_0,a^2e^{itH}\psi_0\rangle
$$

%\end{document}

Since
$$
\langle \psi_0, aa^+ e^{itH}\psi_0 \rangle = \langle \psi_0, (1+a^+a) e^{itH}\psi_0 \rangle
$$
$$
=
\langle \psi_0,  e^{itH}\psi_0 \rangle = f(t)
$$

%\end{document}

and according to (\ref{e5}),
$$
\langle \psi_0, a^2 e^{itH}\psi_0 \rangle = \frac{2}{i\alpha} \frac{df}{dt}
$$
we get:
$$
\frac{df}{dt} = -\frac{\alpha^2}{2\omega }\sin \omega  t \left(\cos \omega  t + \frac{i\beta}{\omega } \sin \omega  t\right)f
$$
$$
-\frac{\alpha^2}{ \omega ^2}\sin^2 \omega  t \frac{df}{dt}
$$
%\end{document}

Therefore, we obtain
$$
\left(1+\frac{\alpha^2}{\omega ^2} \sin^2\omega  t\right)\frac{df}{dt}
$$
$$
=
-\frac{\alpha^2}{2 \omega } \sin \omega  t \left(\cos \omega  t + \frac{i\beta}{\omega } \sin \omega  t\right)f
$$
which is (\ref{eq6}). The initial condition (\ref{df0}) is obvious.

\begin{lemma}\label{lemma6}
If $\omega^2 \ne 0$, then the vacuum expectation of the evolution operator is
\begin{equation}\label{e8}
f(t)=\langle\psi_0,e^{itH(\beta,\omega)}\psi_0\rangle
\end{equation}
%\end{document}

$$
= ({2 e^{-i\beta t}\over(1+{\beta\over\omega}})
e^{-it\omega}+(1-{\beta\over\omega})e^{it\omega})^{1/2}
$$
%\end{equation}
%\end{document}

$$
\Leftrightarrow
f(t)=\langle\psi_0,e^{itH(\beta,\omega)}\psi_0\rangle=
\left(\frac{ e^{-i\beta t}}
{\left( \cos\omega t \ - \ i{\beta\over\omega}\sin\omega t\right)}
\right)^{1/2}
$$
\end{lemma}

%\end{document}

{\bf Proof}. By direct calculations one can check that the function
(\ref{e8}) is the solution of the differential equation (\ref{eq6}) with
the initial condition (\ref{df0}).  In fact:

$$
f(t)=\langle\psi_0,e^{itH(\beta,\omega)}\psi_0\rangle
$$
$$
= \left(\frac{2 e^{-i\beta t}}{\left(1+{\beta\over\omega}\right)
e^{-it\omega}+\left(1-{\beta\over\omega}\right)e^{it\omega}}\right)^{1/2}
$$
$$
f'(t)=
\frac{1}{2}({2 e^{-i\beta t}\over(1+{\beta\over\omega})
e^{-it\omega}+(1-{\beta\over\omega})e^{it\omega}})^{-1/2}
$$
$$
({-i\beta 2 e^{-i\beta t}\over(1+{\beta\over\omega})
e^{-it\omega}+(1-{\beta\over\omega})e^{it\omega}}
$$
%\end{document}

$$-\frac{2 e^{-i\beta t}
(-(1+\beta\/\omega)i\omega
e^{-it\omega}+(1-\beta\/\omega)i\omega e^{it\omega})}
{(1+\beta\/\omega)
e^{-it\omega}+(1-\beta\/\omega)e^{it\omega}
)^2})
$$
%\end{document}

$$
=\frac{1}{2}\frac{1}{f(t)}
\left(- i\beta f(t)^2 + i\omega f(t)^2
\frac{\left(1+{\beta\over\omega}\right)e^{-it\omega}
-\left(1-{\beta\over\omega}\right)e^{it\omega} }
{\left(1+{\beta\over\omega}\right)
e^{-it\omega}+\left(1-{\beta\over\omega}\right)e^{it\omega}}
\right)
$$
%\end{document}
$$
=\frac{i}{2} f(t)
\left(-\beta +
\omega \frac{\left(1+{\beta\over\omega}\right)e^{-it\omega}
-\left(1-{\beta\over\omega}\right) e^{it\omega}}
{\left(1+{\beta\over\omega}\right)e^{-it\omega}
+\left(1-{\beta\over\omega}\right)e^{it\omega} }\right)
$$

%\end{document}

Using
$$
e^{it\omega}= \cos\omega t + i \sin\omega t
$$
one finds that
$$
\left(1+{\beta\over\omega}\right)e^{-it\omega}
+\left(1-{\beta\over\omega}\right)e^{it\omega}
= e^{-it\omega}+{\beta\over\omega}e^{-it\omega}
+e^{it\omega}-{\beta\over\omega}e^{it\omega}
$$
$$
= e^{it\omega} + e^{-it\omega}
+{\beta\over\omega}\left(e^{-it\omega}-e^{it\omega}\right)
$$
$$
= 2  \cos\omega t +{\beta\over\omega}\left(
\cos\omega t -  i \sin\omega t
-\cos\omega t -  i \sin\omega t\right)
$$
$$
= 2\left( \cos\omega t \ - \ i{\beta\over\omega}\sin\omega t\right)
$$
%\end{document}

Similarly
$$
\left(1+{\beta\over\omega}\right)e^{-it\omega}
-\left(1-{\beta\over\omega}\right)e^{it\omega}
=e^{-it\omega}+{\beta\over\omega}e^{-it\omega}
-e^{it\omega}+{\beta\over\omega}e^{it\omega}
$$
$$
=(e^{-it\omega}-e^{it\omega})
+{\beta\over\omega}(e^{-it\omega} +e^{it\omega})
$$
$$
= -2i\sin\omega t +{\beta\over\omega}2\cos\omega t
=  2{\beta\over\omega}
\left(\cos\omega t - i\frac{\omega}{\beta}\sin\omega t\right)
$$
%\end{document}

In conclusion
$$
f'(t)=i/2 f(t) \left(-\beta + \omega
\frac{-i\sin\omega t +{\beta\over\omega}\cos\omega t}
{ \cos\omega t \ - \ i{\beta\over\omega}\sin\omega t}\right)
\Leftrightarrow
$$
$$
(\cos\omega t \ - \ i{\beta\over\omega}\sin\omega t)f'(t)
$$
$$
=\frac{i}{2} f(t) \left(-\beta\cos\omega t \ + \
i{\beta^2\over\omega}\sin\omega t
 - i\omega\sin\omega t +\beta\cos\omega t\right)
\Leftrightarrow
$$

%\end{document}

$$
\Leftrightarrow
=i/2 f(t) \left(
i{\beta^2\over\omega}\sin\omega t
 - i\omega\sin\omega t \right)
\Leftrightarrow
$$
%\end{document}

$$
\Leftrightarrow
=-1/2 f(t) \left(
{\beta^2\over\omega^2}
 - 1\right)\omega\sin\omega t
$$

%\end{document}

Therefore (\ref{e8}) is equivalent to
$$
f(t)=\langle\psi_0,e^{itH(\beta,\omega)}\psi_0\rangle=
\left({ e^{-i\beta t}\over
\cos\omega t \ - \ i{\beta\over\omega}\sin\omega t } \right)^{1/2}
$$

%\end{document}

\begin{lemma}\label{lemma8}
If $\omega^2=0$, then the vacuum expectation of the evolution operator is
\begin{equation}\label{e7}
f(t)=\langle\psi_0,e^{itH}\psi_0\rangle=
\frac{e^{-i\beta t/2}}{\left(1-i\beta t\right)^{1/2}}
\end{equation}
\end{lemma}
{\bf Proof.} Repeating the proofs of Lemmata (\ref{lemma2}), (\ref{lemma3})

%\end{document}

we find that
\begin{enumerate}
\item the
Heisenberg evolutions of $a$, $a^+$ are given by:
$$
a(t)=(1-i\beta t)a-i\beta t a^+\qquad ;\qquad  a^+(t)=(1+i\beta t)a^++i\beta t a
$$
\item $f(t)$ satisfies the following differential equation:
$$
(1+\beta^2 t^2) \frac{df}{dt} = -\frac{\beta^2 t}{2}(i\beta t+1)f
$$
with the initial condition $f(0)=1$.
\end{enumerate}

One can check that (\ref{e7}) is the solution of this equation.

{\bf Remark}. Notice, the the
expression (\ref{e7}) is the limit of the expression (\ref{e8}) as $\omega  \to 0$.

%\end{document}

\section{  The Meixner distributions }

In this section, following \cite{GRIGa}, \cite{GRIGc}
we recall some known facts about Meixner distributions and Meixner classes.
For an historical discussion on their origins we refer to \cite{Meixner}
or to the Appendix of \cite{[AcBou04c_WN calculus and stocastic calculus]}.

\subsection{ The $5$--th Meixner class  }

\begin{definition}
A real valued random variable $X$ is said to have
 Meixner distribution with parameters $a>0$, $b\in(-\pi,\pi)$,
$\mu\in\mathbb{R}$, $\delta>0$ (or to belong to the $5$--th Meixner class)
if its density function is
\begin{equation}
p_X(x;\,a,b,\delta,\mu) =
\frac{(2\cos(b/2))^{2\delta}}
{2a\pi\Gamma(2\delta)}\exp\left(\frac{b(x-\mu)}{a}\right)
\left|\Gamma\left(\delta+\frac{i(x-\mu)}{a}\right)\right|^2
\end{equation}
\end{definition}

%\end{document}

\begin{lemma}
\label{lemma4}
The characteristic function of the Meixner random variable $X$ with
parameters $a,b,\mu,\delta$ is
\begin{equation}\label{meixcf}
E(e^{itX})=
\left({\cos{b\over2}\over\cosh{at-ib\over2}}\right)^{2\delta}e^{i\mu t}
\end{equation}
\end{lemma}
For the proof, see \cite{GRIGa}.

\subsection{ The $4$--th Meixner class: Gamma distributions}

\begin{definition}
A real valued random variable $X$ is said to have the Gamma distribution
with parameters $a,\theta>0$, $\mu \in \mathbb{R}$ (or to belong to the
3--d Meixner class) if its density function is
\begin{equation}\label{e9}
p_{X}(x;\,a,\theta,\mu) =
\frac{(x-\mu)^{a-1}e^{-(x-\mu)/\theta}}{\Gamma(a) \theta^a} \mathbf{1}_{[0,\infty)}
\end{equation}
\end{definition}

\begin{lemma}
\label{lemma7}
The characteristic function of the Gamma random variable $X$ with
parameters $a,\theta,\mu$ is
\begin{equation}
\label{e14}
E(e^{itX})=\frac{e^{-it\mu}}{(1-i\theta t)^a}
\end{equation}
\end{lemma}
\textbf{Proof.}
It is known (for example, see \cite{WolframGamma}) that, denoting ${\cal F}$
the Fourier transform,
\begin{equation}
\label{e13}
{\cal F}\left[\frac{x^{a-1}e^{-x/\theta}}{\Gamma(a) \theta^a}
\chi_{[0,\infty)}(x)\right](t)=(1-i\theta t)^{-a}
\end{equation}
Moreover, for any $\phi$, such that ${\cal F}[\phi]$ exists,
\begin{equation}
\label{e12}
{\cal F}[\phi(x-\mu)](t) = e^{-it\mu} {\cal F}[\phi(x)](t)
\end{equation}
Combining (\ref{e13}) and (\ref{e12}) we have (\ref{e14}).

%\end{document}

\subsection{ The $3$--d Meixner class: Negative Binomial (Pascal) distributions}

\begin{definition}
A real valued random variable $X$ is said to have Negative Binomial
(or Pascal) distribution with parameters $0<p<1$, $r \ne 0$,
$\mu \in \mathbb{R}$, and $d \ne 0$ (or to belong to the 4--th Meixner class)
if its probability density function is given by
$$
P(x;\, r,p,\mu,d) =
\sum_{n=0}^{\infty} \left(
\begin{array}{c} r \\ n\end{array}\right)p^r (1-p)^n \delta(x-nd-\mu)
$$
where, by definition, for any $r \in \mathbb{R}$, $k \in \Bbb N \cup \{0\}$:
$$
\left(\begin{array}{c} r\\ k\end{array}\right):=
\frac{r(r+1)\dots(r+k-1)}{k!}\qquad ;\qquad k\ge 1;\mbox{  and  }
\left(\begin{array}{c} r\\ 0\end{array}\right):=1
$$
\end{definition}

{\bf Remark}.
Recall, for completemess, that the $2$--d Meixner class is the Poisson
and the $1$--st one is the Gaussian.

%\end{document}

\section{   Meixner distributions and quadratic Bose Hamiltonians  }

\begin{theorem}
Suppose that the condition
\begin{equation}\label{nc}
\det h = -\omega^2  = \alpha^2 - \beta^2 > 0
\end{equation}
is satisfied. Then the vacuum expectation
$f(t)=\langle\psi_0,e^{itH}\psi_0\rangle$,
with $H$ given by (\ref{e2}), (\ref{e2real}), is the characteristic function of
the Meixner type $5$ random variable with parameters
\begin{equation}\label{idpar5}
a=2|\omega| \quad , \quad b=2i\log\frac{\omega+\beta}{i\alpha} \quad , \quad
\mu=-\frac{\beta}{2} \quad , \quad \delta=\frac{1}{4}
\end{equation}
More explicitly:
\begin{equation}\label{e25a}
\left(
\frac{2e^{-i\beta t}}{\left(1+\frac{\beta}{\omega}\right)e^{-it\omega}+\left(1-\frac{\beta}{\omega}\right)e^{it\omega}} \right)^{1/2}
=e^{it\beta/2}\left(\frac{\cos \frac{b}{2}}
{\cosh \frac{at-ib}{2} }\right)^{2\cdot\frac{1}{4}}
\end{equation}
\end{theorem}
\textbf{Proof.}
The idea of the proof is to fit the parameters $a$, $b$, $\mu$ and
$\delta$, given by (\ref{idpar5}) with those of the characteristic function given by
Lemma {\ref{lemma4}}.
If $\omega^2<0$, then $\omega = -i |\omega|$. Hence,
\begin{equation}\label{e18}
e^{{at}/{2}} = e^{|\omega|t} = e^{-i\omega t}
\end{equation}
Using the definition of $\omega$, we have
\begin{equation}\label{e20}
(\omega+\beta)(\omega-\beta)=\omega^2-\beta^2 = -\alpha^2 = (i\alpha )^2
\Leftrightarrow
{\omega+\beta \over i\alpha}  = {i\alpha \over \omega-\beta }
\end{equation}
In particular from this and (\ref{idpar5}) it follows that:
\begin{equation}\label{e22}
e^{-{ib}/{2}} = e^{\log \frac{\omega+\beta}{i\alpha}}
= \frac{\omega+\beta}{i\alpha}
= {i\alpha \over \omega-\beta } \Leftrightarrow
e^{{ib}/{2}} =  \frac{\omega-\beta}{i\alpha}
\end{equation}
Using (\ref{e22}) and recalling that $\omega$ is purely
imaginary and $\beta, \alpha$ are real, we find:
\begin{equation}\label{e23}
\cos \frac{b}{2} = Re\left( e^{-{ib}/{2}} \right) = \frac{\omega}{i\alpha}
\end{equation}

%\end{document}

From (\ref{e18},\ref{e22}) we have:
$$
\cosh \frac{at-ib}{2} =
 \frac{1}{2}\left(e^{at/2}e^{-ib/2}+e^{-at/2}e^{ib/2}\right)
 $$
 $$
 =
\frac{1}{2}\left(e^{-i\omega t}\,\frac{\omega+\beta}{i\alpha}+
e^{i\omega t}\,\frac{\omega-\beta}{i\alpha}\right)
$$
\begin{equation}\label{e24}
= \frac{1}{2}\frac{\omega}{i\alpha}\,
\left((1 +\frac{\beta}{\omega})e^{-it\omega }\, + \,
(1 -\frac{\beta}{\omega})e^{it\omega }\right)
\end{equation}
$$
= \frac{1}{2}\cos \frac{b}{2}
\left((1 +\frac{\beta}{\omega})e^{-it\omega }\, + \,
(1 -\frac{\beta}{\omega})e^{it\omega }\right)
$$
%\end{equation}

%\end{document}

Replacing (\ref{e23}) and (\ref{e24}) in the characteristic function
(\ref{e8}), we obtain:
$$
e^{i(\beta/2)t}\left(\cos \frac{b}{2}\right)^{2\cdot\frac{1}{4}}
\left(\cosh \frac{at-ib}{2} \right)^{-2\cdot\frac{1}{4}} =
$$
$$
=
e^{-i\beta t/2} \left(\frac{\omega}{i\alpha}\right)^{-1/2}
\left(\frac{1}{2}\right)^{-1/2} \left(\frac{\omega+\beta}{i\alpha}\,e^{-i\omega t}+\frac{\omega-\beta}{i\alpha}\,e^{i\omega t}\right)^{-1/2}
$$
\begin{equation}\label{e25}
=\frac{\sqrt{2}e^{-i\beta t/2}}{\left(\left(1+\frac{\beta}{\omega}\right)e^{-it\omega}+\left(1-\frac{\beta}{\omega}\right)e^{it\omega}\right)^{1/2}}
\end{equation}
which is the vacuum expectation given by Lemma {\ref{lemma6}}.

%\end{document}

\begin{definition}
We say that two characteristic functions $\phi_1(t)$ and $\phi_2(t)$ are equal
up to simple transformations if $\phi_1(t)$ can be transformed into $\phi_2(t)$
by applying a finite sequence of the following transformations:

"time shift"
$$\phi(t) \to \phi(t)e^{imt}\qquad ;\qquad  m\in \mathbb{R}\,,$$
"independent copying"
$$\phi(t) \to \phi(t)^q\qquad ;\qquad  q>0\,,$$
"time rescaling"
$$
\phi(t) \to \phi(kt)\qquad ;\qquad  k \ne 0\,.
$$
\end{definition}
\begin{proposition}\label{oldRemark1.}
Up to simple transformations the characteristic function of any Meixner
random variable of type V, can be reduced to a vacuum expectation of the
form (\ref{e8}) for some homogeneous quadratic Bose Hamiltonian
$H$ satisfying (\ref{nc}).
\end{proposition}
\textbf{Proof.}
Consider the Meixner type V random variable $X$
with parameters $(a,b,\delta,\mu)$ and put:
\begin{equation}\label{Z1}
\alpha = \frac{a}{{2}} (\cos \frac{b}{2})^{-1}
\end{equation}
\begin{equation}\label{Z2}
\beta = \frac{a}{{2}} \tan \frac{b}{2}
\end{equation}
Then the vacuum expectation of the quadratic Hamiltonian, specified by
these parameters, is the
characteristic function $\phi_{a',b',\delta',\mu'}(t)$ of Meixner distribution with
parameters $a'$, $b'$, $\delta'$, and $\mu'$, where:
$$
a' = 2|\omega|  =
2\left|\sqrt{\left(\frac{a}{2}\tan \frac{b}{2}\right)^2-
\left(\frac{a}{2} (\cos \frac{b}{2})^{-1}\right)^2}\,\right| = $$
$$
a\left|\sqrt{\tan^2 \frac{b}{2} - (\cos \frac{b}{2})^{-2}}\,\right| =
a \left| \sqrt{\frac{\sin^2(b/2)-1}{\cos^2(b/2)}} \,\right| =
a|i| = a
$$
$$
e^{\frac{-ib'}{2}} = \frac{\omega+\beta}{i\alpha} =
\frac{\frac{a}{2}\sqrt{\tan^2\frac{b}{2}-(\cos\frac{b}{2})^{-2}} +
\frac{a}{2}\tan \frac{b}{2}}{i\frac{a}{a}(\cos\frac{b}{2})^{-1}} =
$$
$$
= \frac{i+ \tan \frac{b}{2}}{i^{-1}\cos\frac{b}{2})^{-1}} =
\cos\frac{b}{2} - i \sin\frac{b}{2} = e^{\frac{ib}{2}}
$$
and, therefore, $b'=b$, $\mu'=\frac{1}{2}b$ and $\delta'=\frac{1}{4}$.
For any $\delta,\mu$ we have:
$$
\phi_{a,b,\delta,\mu}(t) =
e^{i\mu t} (e^{-i \mu' t} \phi_{a,b,1/4,\mu'}(t))^{4\delta}
$$
This proves the statement.

%\end{document}

\begin{theorem}
Suppose that
\begin{equation}\label{det0}
\det h = -\omega^2 =0
\end{equation}
Then the vacuum expectation ${f(t)=\langle\psi_0,e^{itH}\psi_0\rangle}$ is the
characteristic function of the Gamma distribution with parameters
$$
a=1/2 \quad , \quad \theta=\beta \quad , \quad \mu=\beta/2
$$
\end{theorem}

\textbf{Proof.}
This follows from Lemmata (\ref{lemma8}) and (\ref{lemma7}).

\textbf{Remark 2.} It is easy to verify that the characteristic function
of any Gamma distributed random variable, up to simple transformations, is the
vacuum expectation of some Bose homogeneous quadratic Hamiltonian
satisfying (\ref{det0}).

\begin{theorem}\label{theoremY}
Suppose
\begin{equation}\label{det<0}
\det h = -\omega^2 < 0
\end{equation}
Then the vacuum expectation
${f(t)=\langle\psi_0,e^{itH}\psi_0\rangle}$ is the characteristic
function of the Negative Binomial distribution with parameters
\begin{equation}\label{parnegbin}
r=1/2\quad , \quad p=\frac{2\omega}{\omega+\beta}\quad , \quad
{\mu=\frac{\beta-\omega}{2}}\quad , \quad d=-2\omega
\end{equation}
\end{theorem}

\textbf{Proof.} From Lemma \ref{lemma6} we have:
$$
f(t)=\left({2 e^{-i\beta
t}\over\left(1+{\beta\over\omega}\right)
e^{-it\omega}+\left(1-{\beta\over\omega}\right)e^{it\omega}}\right)^{1/2} =
\sqrt{2}e^{-i\beta t/2} \left(\frac{\omega+\beta}{\omega}\,e^{-i \omega t} -
\frac{\beta-\omega}{\omega}\, e^{i \omega t}\right)^{-1/2}
$$
$$
=\sqrt{2}e^{-it\frac{\beta-\omega}{2}}\, \left(\frac{\omega}{\omega+\beta}\right)^{1/2}
\left(1-\frac{\beta-\omega}{\beta+\omega} e^{2i\omega t}\right)^{-1/2}
$$
Note that
$$
1-\frac{2\omega}{\beta+\omega} = \frac{\beta-\omega}{\beta+\omega}
$$
therefore, defining $p$, $d$, and $\mu$ through (\ref{parnegbin}) we have:
$$
f(t)=p^{1/2} e^{-i \mu t} \frac{1}{\sqrt{1-(1-p)e^{-i d t}}}
$$
whose series expansion around $x=0$ is:
\begin{equation}
\label{W1}
\frac{1}{\sqrt{1-x}} = \sum_{n=0}^{\infty} c_n x^n
\end{equation}
and the coefficients $c_n$ are given by Newton's binomial:
$$
c_n=\frac{1}{n!} \left.\frac{d^n}{dx^n} \frac{1}{\sqrt{1-x}}\right|_{x=0} =
\frac{1}{n!} \cdot \frac{1}{2} \cdot \frac{3}{2}\cdot \dots \cdot \frac{2n-1}{2} =
\left(\begin{array}{c} 1/2 \\ n \end{array}\right)\qquad ; n\ge 1
$$
and $c_0=1$. Using this expansion, we get the series expansion:
$$
f(t)=K e^{-i\mu t} \sum_{n=0}^{\infty} c_n (1-p)^n e^{-ind t}
$$
which, since $|e^{-ind t}|=1$, is uniformly convergent for any $t$
and for any $p$ in a bounded set.
In our case since $\omega^2>0$, (\ref{parnegbin}) implies that
$-1<1-p=\frac{\beta-\omega}{\beta+\omega}<1$, i.e.
$|1-p|<1$ and the series converges uniformly for all $t$.
The inverse Fourier transform of $f$ is
$$
F^{-1}[f(t)](x) = \frac{1}{2\pi}\int_{-\infty}^{\infty} e^{ixt} f(t)\, dt =
\frac{p^{1/2}}{2\pi}\int_{-\infty}^{\infty}e^{i\left(x-\mu\right)t}
\sum_{n=0}^{\infty} c_n (1-p)^n e^{-ind t}
$$
Hence,
$$
F^{-1}[f(t)](x) =
\frac{p^{1/2}}{2\pi} \sum_{n=0}^{\infty} c_n (1-p)^n \int_{-\infty}^{\infty}
e^{i\left(x-\mu-nd\right)t}\, dt
$$
$$
=
\sum_{n=0}^{\infty} c_n p^{-1/2} (1-p)^n \delta(x-\mu-nd) =
=\sum_{n=0}^{\infty} \left(\begin{array}{c} 1/2\\ n \end{array}\right) p^{-1/2} (1-p)^n \delta(x-\mu-nd)
$$
which is density function of the Pascal distribution with $r=1/2$.

%\end{document}

\begin{proposition}\label{oldRemark3.}
The characteristic function of any Negative Binomial
distribution is, up to simple transformations, the vacuum expectation of
some $1$-dimensional homogeneous quadratic Bose Hamiltonian satisfying (\ref{det0}).
\end{proposition}

\textbf{Proof.}
Given $0<p<1$, $r\ne 0$, $\mu \in \mathbb{R}$, and $d \ne 0$,
let the parameters $\alpha$ and $\beta$ be:
\begin{equation}
\label{Y1}
\alpha = \sqrt{1-\left(\frac{p}{2-p}\right)^2}\qquad ;\qquad  \beta = 1
\end{equation}
Then the vacuum expectation of the quadratic Bose Hamiltonian, specified by these
$\alpha$ and $\beta$ is:
$$
f(t) = {p'}^{1/2} e^{-i \mu' t} \left(1-(1-{p'})e^{-id't}\right)^{1/2}
$$
where
$$
\omega = \sqrt{\beta^2-\alpha^2} =
\sqrt{1-\left(1-\left(\frac{p}{2-p}\right)^2\right)} = \frac{p}{2-p}
$$
$$
p' = \frac{2\omega}{\omega+\beta} =
\frac{p}{2-p}\left(\frac{p}{2-p}+1\right)^{-1} =
\frac{2p}{2-p} \left(\frac{2}{2-p}\right)^{-1} = p
$$
$$
\mu'=\frac{1-\omega}{\omega}\qquad ;\qquad d'=-2\omega
$$
Let us apply the following (simple) transformation:
$$
f(t) \to e^{i\mu t }(e^{-i\mu'\lambda t}f(\lambda t))^{2r}
$$
where $\lambda = \frac{d}{d'}$. We have:
$$
f(t) \to f_2(t) = p^r \left(1-(1-p)e^{idt}\right)^{-r} e^{i\mu t }
$$
Applying the Newton binomial formula:
\begin{equation}\label{Y2}
(1-x)^{-a}=\sum_{n=0}^{\infty} \left(\begin{array}{c} a \\ n \end{array}\right) x^n
\end{equation}
and repeating computations of Theorem \ref{theoremY}
one can find that $f_2(t)$ is the characteristic function of the distribution
$$
\sum_{n=0}^{\infty}
\left(
\begin{array}{c}
a \\
n
\end{array}
\right)
p^r (1-p)^n \delta(x-dn-\mu)
$$
Which is the negative Binomial distribution with parameters
$p$, $r$, $\mu$, and $d$.

We summarize Remarks 1-3 in the following theorem:
\begin{theorem}
The characteristic function of any Meixner type III,IV,V distribution is,
up to a simple transformation,  the vacuum expectation of some
$1$--dimensional homogeneous quadratic Bose Hamiltonian.
\end{theorem}

%\end{document}

\section{$n$-dimensional case}
\label{n-dim-cse}

Let $a_i$, $a_i^+$, $i=1,2,\dots, n$ be Bose annihilation and creation operators,
satisfying CCR:
$$
[a_i,a_j]=[a_i^+,a_j^+]=0\qquad ;\qquad [a_i,a^+_j]=\delta_{i,j}\qquad ;\qquad  i,j=1,2,\dots n
$$
Because of the commutativity of the creators (annihilators),
the most general Hermitean quadratic expression in the $a_i^\pm$
%(Hamiltonian)
which is real homogeneous of degree $2$ is:
\begin{equation}\label{e21a}
H_{A,C} =\sum_{i,j=1}^{n} A_{ij} a^+_i a^+_j +
\sum_{i,j=1}^{n} \overline A_{ij} a_i a_j +
\sum_{i,j=1}^{n} C_{ij} a^+_i a_j
\end{equation}
where $A_{ij}, C_{ij} \in \mathbb{C}$.
The Hermiteanity condition for $H_{A,C} $ and the mutual commutativity of
creators (resp. annihilators) imply that
\begin{equation}\label{e21b}
H_{A,C} =H_{A,C}^*=\sum_{i,j=1}^{n} \overline A_{ij}a_ja_i +
\sum_{i,j=1}^{n} A_{ij} a^+_ja^+_i  +
\sum_{i,j=1}^{n} \overline C_{ij} a^+_ja_i
\end{equation}
$$
=\sum_{i,j=1}^{n} A_{ji} a^+_i a^+_j +
\sum_{i,j=1}^{n}  \overline A_{ji} a_i a_j +
\sum_{i,j=1}^{n} \overline C_{ji} a^+_i a_j
$$
Therefore
\begin{equation}\label{e21c}
 H_{A,C} =\frac{1}{2}(H_{A,C} +H_{A,C}^*) =
\end{equation}
$$
=\sum_{i,j=1}^{n} \frac{1}{2}(A_{ij}+ A_{ji}) a^+_i a^+_j +
\sum_{i,j=1}^{n}  \frac{1}{2}(\overline A_{ij}+ \overline A_{ji}) a_i a_j +
\sum_{i,j=1}^{n} \frac{1}{2}(C_{ij}
$$
$$
+ \overline C_{ji})  a^+_i a_j
$$
Therefore one can suppose that
$$
A_{ij} = A_{ji} \qquad;\qquad
C_{ij}= \overline C_{ji}
$$
i.e. that the $n\times n$ matrices
$A:=(A_{ij})$ and $C:=(C_{ij})$ are respectively symmetric
and Hermitean:
$$
A= A^T \qquad ; \qquad C= C^*
$$
Denote the $n$-component
vectors $(a_i)$ and $(a^+_i)$ by $a$ and $a^*$.
Then, up to an additive constant, one can rewrite the Hamiltonian
(\ref{e21a}) in matrix form:
\begin{equation}\label{e21}
H_{A,C}=\frac{1}{2}\left( a^* A a^* + a A a\right) + a^* C a =
{1\over2}\,(a^+,a)
 \pmatrix{C  &A\cr  \bar A& \bar C\cr}\pmatrix{a\cr a^+\cr}=H
\end{equation}

\begin{theorem}\label{struceith}
There exist $n \times n$ matrices $\Phi(t)$ and $\Psi(t)$ such that:
$$
   \pmatrix{\Phi(t) & \Psi(t) \cr
  \bar \Psi(t) &  \bar \Phi(t) \cr}
=
\exp \left( it{1\over2}\,
 \pmatrix{C  &A\cr \overline A& \overline C\cr} \right)
$$

\end{theorem}

{\bf Proof}. This result was proved by Friedrichs and extended by
Berezin (\cite{Berezin} pg. 122).\\

%\end{document}

We denote $\psi_0$ the vacuum vector, characterized by $a_j\psi_0=0$
for all $j=1,2,\dots n$.

\begin{theorem}\label{theoremfriedber}
$$
\langle \psi_0,\ e^{itH_{A,C}} \psi_0 \rangle
=\frac{1}{\sqrt{\det \Phi(t) e^{iCt} }}
=\det (\Phi(t))^{-1/2} \det (e^{iCt})^{-1/2}
$$
\end{theorem}
{\bf Proof}. This result was proved by Friedrichs and extended by
Berezin (\cite{Berezin} pg. 122).
\begin{lemma}\label{lemma12}
$\Phi(t)$ satisfies the following equation:
\begin{equation}\label{e12_00}
A^{-1}\ddot \Phi - i[C,A^{-1}]\dot \Phi + (CA^{-1}C - A) \Phi = 0
\end{equation}
with the initial conditions $\Phi(0)=1$, $\dot \Phi(0) = - iC$.
\end{lemma}
\textbf{Proof.}
From the definition it follows that
\begin{equation}
\label{e12_005}
\frac{d}{dt}
\left(%
\begin{array}{cc}
  \Phi & \Psi \\
  \bar \Psi &  \bar \Phi \\
\end{array}%
\right)=
i
\left(%
\begin{array}{cc}
  -C & -A \\
  A &  C \\
\end{array}%
\right)
\left(%
\begin{array}{cc}
  \Phi & \Psi \\
  \bar \Psi &  \bar \Phi \\
\end{array}%
\right)
\end{equation}
In particular,
\begin{equation}
\label{e12_01}
\dot \Psi = - i C \Psi - i A \bar\Phi
\end{equation}
\begin{equation}
\label{e12_02}
\dot {\bar
 \Phi} = i A \Psi + i C \bar \Phi
\end{equation}
Let us express $\Psi$ using (\ref{e12_02}):
\begin{equation}
\label{e12_03}
\Psi = \frac{1}{i} A^{-1} \left( \dot {\bar
 \Phi} - i C \bar \Phi \right)
\end{equation}
Substituting this for $\Psi$ into (\ref{e12_01}) we obtain:
$$
\frac{d}{dt}\left(\frac{1}{i} A^{-1} \left( \dot {\bar \Phi} - i C \bar \Phi \right)\right) =
-i C \left(\frac{1}{i} A^{-1} \left( \dot {\bar \Phi} - i C \bar \Phi \right)\right) - i A \bar \Phi
$$
$$
-iA^{-1} \ddot{\bar \Phi}+(CA^{-1}-A^{-1}C)\dot{\bar{\Phi}}+i\left(A-CA^{-1}C\right)\bar \Phi = 0
$$
Multiplying by $i$ and taking the adjoint, we have:
$$
A^{-1} \ddot{\Phi}+i(CA^{-1}-A^{-1}C)\dot{{\Phi}}+\left(CA^{-1}C-A\right)\Phi = 0
$$
which is (\ref{e12_00}). The initial conditions are clear.

\subsection{ Vector valued Meixner random variables }

Recall that a random variable $X$, with values in a finite dimensional real
vector space $V$ (identified to its dual space) is defined by a linear map
$$
X:V\to \ \hbox{real valued random variables}
$$
An homogeneous quadratic Bose Hamiltonian $H$ of the form (\ref{e21a}),
is uniquely determined by a pair $(A,C)$ where $A$ is a symmetric
complex $d\times d$ matrix and $C$ a complex self--adjoint matrix.
The set of such pairs has a natural structure of finite dimensional
real vector space, that we denote $V_d$, and the map
$$
(A,C)\mapsto H_{A,C} \ \in \ \{\hbox{self--adjoint operators}\}
$$
is clearly real linear. It is known that each $H_{A,C}$ can be identified to
a real valued classical random variable with respect to the vacuum vector.
\begin{definition}\label{dfvectvalmeixn}
In the above notations,
a  $V_d$--valued random variable $X$ is called of Meixner type
if for each $(A,C)\in V_d$ the characteristic function of
$X_{(A,C)}$ coincides with the vacuum characteristic function of $H_{A,C}$:
$$
E\left(e^{itX_{(A,C)}}\right)=\langle \psi_0,\ e^{itH_{A,C}} \psi_0 \rangle
$$
where $H_{A,C}$ denotes the homogeneous quadratic Fock Bose Hamiltonian
with $d$--degrees of freedom, $ \psi_0$ the corresponding vacuum vector
and $E$ the expectation with respect to the random variable $X_{(A,C)}$.
\end{definition}
{\bf Remark}. From Theorem (\ref{struceith}) we then know that Definition
(\ref{dfvectvalmeixn}) is equivlent to require that
\begin{equation}\label{charfunvecmeix}
E\left(e^{itX_{(A,C)}}\right)=\det(\Phi (t)\cdot e^{iCt)})^{-1/2}
\end{equation}
where $\Phi (t)$ is the solution of equation (\ref{e12_00}).

\subsection{ The case of commuting $A$ and $C$  }

\begin{theorem}\label{commAC-impl-prod-Meixn}{\rm
Let the $d\times d$ Hermitean complex matrices $A$, $C$, $\Phi (t)$
be as in Theorem (\ref{struceith}) and suppose that
\begin{equation}\label{comm-AC-1}
[A,C]=0
\end{equation}
or, equivalently, that $A$ and $C$ can be simultaneously
diagonalized by an orthogonal transformation $Q$:
\begin{equation}\label{eig-val-A}
Q^T A Q = \Lambda_A = diag [\alpha_1,\alpha_2,\dots,\alpha_n]
\end{equation}
\begin{equation}\label{eig-val-C}
Q^T C Q = \Lambda_C = diag[\beta_1,\beta_2,\dots,\beta_n]
\end{equation}
Then
\begin{equation}\label{e13_00}
\left(\det \Phi e^{iCt}\right)^{-1/2} =
\left(
\frac{ e^{-it (\beta_1+\beta_2+\dots+\beta_n)}}{
\prod_{j=1}^{n} \left(\cos \omega_j t - \frac{i \beta_j}{\omega_j}
\sin \omega_j t \right) }\right)^{1/2}=
\end{equation}
$$
=\prod_{j=1}^{n} \left(
\frac{ e^{-it \beta_j}}{
\cos \omega_j t - \frac{i \beta_j}{\omega_j}
\sin \omega_j t  }\right)^{1/2}
$$
where
\begin{equation}
\label{e13_01}
\omega_i^2 = \beta_i^2-\alpha_i^2
\end{equation}
In particular the vacuum distribution of the symmetric operator
$H_{A,C}$, given by (\ref{e21}) is a product of Meixner distributions.
}\end{theorem}
\textbf{Proof.}
Since $A$ and $C$ are commuting symmetric real matrices with
eigenvalues $\alpha_i, \beta_i \in \mathbb{R}$ respectively, then
$$
\left(\det \Phi e^{iCt}\right)^{-1/2}
=\left(\det \Phi\right)^{-1/2}\left(\det e^{iCt}\right)^{-1/2}
$$
Notice that $\det Q = 1$. Hence,
\begin{equation}\label{e13_02}
\det e^{iCt}
= \det e^{it Q^T \Lambda_C Q} = \det \left(Q^T e^{it\Lambda_C} Q \right)
=\det e^{it\Lambda_C} = \prod_{j=1}^{n} e^{it\beta_j}
= e^{it(\beta_1+\beta_2+\dots+\beta_n)}
\end{equation}
Secondly, denote $\Phi'=Q^T \Phi Q$.
Then Eq. (\ref{e12_00}) in terms of $\Phi'$  becomes
$$
\ddot \Phi' + (\Lambda_C^2 - \Lambda_A^2) \Phi' = 0
$$
Denote the components of the $\Phi'$ matrix by
$\phi_{ij}$, $i,j=1,2,\dots,n$.
The equation for each component is independent.
For the off-diagonal components we have:
$$
\frac{d^2}{dt^2}{\phi}_{ij} = 0
\qquad ;\qquad
\phi_{ij}(0)=0\qquad ;\qquad  \frac{d}{dt}\phi_{ij}(0) = 0\qquad ;\qquad  i \ne j
$$
With the obvious solution $\phi_{ij}(t)=0$. For the diagonal components we have:
$$
\frac{d^2}{dt^2}{\phi}_{ii}+ \omega_i^2 {\phi}_{ii} = 0\qquad ;\qquad
\phi_{ii}(0) = 1\qquad ;\qquad  \dot\phi_{ii}(0) = -i \beta_i
$$
where $\omega_i$ is given by (\ref{e13_01}).
Solving this equation, we have:
$$
\phi_{ii}(t)
= \cos \omega_i t - \frac{i \beta_i}{\omega_i} \sin\omega_i t
$$
Thus, we see that $\Phi'$ is a diagonal matrix and
\begin{equation}
\label{e13_04}
\det \Phi' = \prod_{j=1}^{n} \phi_{jj}(t)
= \prod_{j=1}^{n} \left(\cos \omega t
- \frac{i \beta_i}{\omega_i} \sin \omega_i t \right)
\end{equation}
Finally, note that
\begin{equation}
\label{e13_05}
\det \Phi' = \det Q^T \Phi Q = \det \Phi
\end{equation}
Combining (\ref{e13_02}), (\ref{e13_04}), and (\ref{e13_05})
we obtain  (\ref{e13_00}).

\subsection{  The $1$--dimensional case}

Consider the
simplest quantum model with the Hilbert space
${\cal H}=L^2(\Bbb R)$
and the pair of creation and annihilation operators $a^+$ and $a$ satisfying
$$
[a,a^+]=1
$$
Denote $\psi_0\in{\cal H}$ the vector (vacuum)
such that
$$
a\psi_0=0
$$
We will determine the explicit form of the right
hand side of (\ref{theoremfriedber}) when $n=1$.
In this case $A$ and $C$ are real numbers and
$\Phi (t)$ is a complex valued function.
\begin{theorem}
If $n=1$ then the following identity holds
\begin{equation}\label{(9)}
(\Phi (t)e^{iCt})^{-1/2}=
\sqrt{2}{e^{-iCt/2}\over\left[\left(1+{C\over\omega}\right)
e^{-i\omega t}+\left(1-{C\over\omega}\right)e^{i\omega t}\right]^{1/2}}
\end{equation}
where
\begin{equation}\label{9a}
\omega=\sqrt{C-|A|^2}
\end{equation}
\end{theorem}

\noindent{\bf Proof}.
Let us verify the equality (\ref{(9)}). From Theorem (\ref{struceith}) one gets
\begin{equation}
\pmatrix{
\dot \Phi &\dot \Psi\cr
\dot{\overline \Psi}&\dot{\overline \Phi }\cr}=ih\pmatrix{
               \Phi  &                \Psi\cr
\overline \Psi&\overline \Phi \cr}\label{(10)}
\end{equation}
or
\begin{equation}
\dot \Phi =-iC\Phi -iA\overline \Psi\label{(11)}
\end{equation}
$$\dot \Psi=-iC\Psi-iA\overline \Phi $$
The initial boundary conditions are
\begin{equation}
\Phi (0)=1\ ,\quad\dot \Phi (0)=-iC\ ;\quad \Psi(0)=0\ ,\quad\dot \Psi(0)=-iA\label{(12)}
\end{equation}
From this one can show that $\Phi (t)$ has the form
$$\Phi (t)={1\over2}\,\left[\left(1+{C\over\omega}\right)e^{-i\omega t}+\left(1-{C\over
\omega}\right)e^{i\omega t}\right]$$
and this proves the relation (\ref{(9)}).

{\bf Acknowledgments}
This work was supported by the grant of the Russian Science Foundation
 RSF 14-11-00687.

\end{document}